\documentclass[amsmath,amssymb,prb,preprintnumbers,superscriptaddress, twocolumn,showpacs]{revtex4}
\usepackage{graphicx}
\usepackage{amsmath}
\usepackage{amssymb}
\usepackage{bm}         
\usepackage[squaren, thinqspace]{SIunits}
\usepackage{xspace}
\usepackage{setspace}
\usepackage{textcomp}
\usepackage[usenames]{color} 
\usepackage{soul}
\usepackage{ulem}       

\newcommand{\lapprox}{{\scriptscriptstyle\stackrel{<}{\sim}}}

\newif\ifcom
\newif\ifdel

\comtrue               %

\deltrue               %

\begin{document}


\title{Edge superconductivity in Nb thin film microbridges revealed by integral and spatially resolved electric transport}

\author{R.~Werner}
\affiliation{Physikalisches Institut -- Experimentalphysik II and Center for Collective Quantum Phenomena in LISA$^+$,
Universit\"{a}t T\"{u}bingen, Auf der Morgenstelle 14, 72076
T\"{u}bingen, Germany}
\author{A.~Yu.~Aladyshkin}
\affiliation{Institute for Physics of Microstructures RAS, 603950,
Nizhny Novgorod, GSP-105, Russia}
\author{I.~M.~Nefedov}
\affiliation{Institute for Physics of Microstructures RAS, 603950,
Nizhny Novgorod, GSP-105, Russia}
\author{A.~V.~Putilov}
\affiliation{Institute for Physics of Microstructures RAS, 603950,
Nizhny Novgorod, GSP-105, Russia}
\author{M.~Kemmler}
\affiliation{Physikalisches Institut -- Experimentalphysik II and Center for Collective Quantum Phenomena in LISA$^+$,
Universit\"{a}t T\"{u}bingen, Auf der Morgenstelle 14, 72076
T\"{u}bingen, Germany}
\author{D.~Bothner}
\affiliation{Physikalisches Institut -- Experimentalphysik II and Center for Collective Quantum Phenomena in LISA$^+$,
Universit\"{a}t T\"{u}bingen, Auf der Morgenstelle 14, 72076
T\"{u}bingen, Germany}
\author{A.~Loerincz}
\affiliation{Institut f\"{u}r Mikro- und Nanoelektronische Systeme, Karlsruher Institut f\"{u}r Technologie, Hertzstra{\ss}e 16, D-76187 Karlsruhe, Germany}
\author{K.~Ilin}
\affiliation{Institut f\"{u}r Mikro- und Nanoelektronische Systeme, Karlsruher Institut f\"{u}r Technologie, Hertzstra{\ss}e 16, D-76187 Karlsruhe, Germany}
\author{M.~Siegel}
\affiliation{Institut f\"{u}r Mikro- und Nanoelektronische Systeme, Karlsruher Institut f\"{u}r Technologie, Hertzstra{\ss}e 16, D-76187 Karlsruhe, Germany}
\author{R.~Kleiner}
\affiliation{Physikalisches Institut -- Experimentalphysik II and Center for Collective Quantum Phenomena in LISA$^+$,
Universit\"{a}t T\"{u}bingen, Auf der Morgenstelle 14, 72076
T\"{u}bingen, Germany}
\author{D.~Koelle}
\affiliation{Physikalisches Institut -- Experimentalphysik II and Center for Collective Quantum Phenomena in LISA$^+$,
Universit\"{a}t T\"{u}bingen, Auf der Morgenstelle 14, 72076
T\"{u}bingen, Germany}

\date{\today}
\begin{abstract}
%
The resistance $R$ vs perpendicular external magnetic field $H$ was measured for superconducting Nb thin--film microbridges with and without microholes [antidots (ADs)].
Well below the transition temperature, integral $R(H)$ measurements of the resistive transition to the normal state on the plain bridge show two distinct regions, which can be identified as bulk and edge superconductivity, respectively.
The latter case appears when bulk superconductivity becomes suppressed at the upper critical field $H_{c2}$ and below the critical field of edge superconductivity $H_{c3}\approx 1.7\, H_{c2}$.
The presence of additional edges in the AD bridge leads to a different shape of the $R(H)$ curves.
We used low-temperature scanning laser microscopy (LTSLM) to visualize the current distribution in the plain and AD bridge upon sweeping $H$.
While the plain bridge shows a dominant LTSLM signal at its edges for $H > H_{c2}$ the AD bridge also gives a signal from the inner parts of the bridge due to the additional edge states around the ADs.
LTSLM reveals an asymmetry in the current distribution between left and right edges, which confirms theoretical predictions.
Furthermore, the experimental results are in good agreement with our numerical simulations (based on the time-dependent Ginzburg--Landau model) yielding the spatial distribution of the order parameter and  current density for different bias currents and $H$ values.
\end{abstract}
%



\maketitle

\section{Introduction}
The concept of  localized superconductivity in bulk superconductors was introduced in 1963 by Saint-James and de Gennes \cite{Saint-James63}.
They demonstrated that superconductivity in a semi--infinite sample with an ideal flat surface in the presence of an external magnetic field $\bm H$ (with amplitude $H$) parallel to its surface can survive in a thin surface layer, even above the upper critical field $H_{c2}$, when bulk superconductivity is completely suppressed.
Based on the phenomenological Ginzburg--Landau theory, the critical field $H_{c3}$ for surface superconductivity, localized near  superconductor/vacuum or superconductor/insulator interfaces, can be calculated as \cite{Abrikosov88,Tinkham96}
\begin{equation}
1.695\,H_{c2}\simeq H_{c3} =H_{c3}^{(0)} \left(1-T/T_{c0}\right),
\label{eq:Hc3}
\end{equation}
where $H_{c3}^{(0)}$ is the upper critical field for surface superconductivity at temperature $T=0$, and $T_{c0}$ is the superconducting critical temperature for $H=0$.
This theory predicts that in the regime of the surface superconductivity the order parameter wave function $\Psi$ decays exponentially with increasing distance from the surface on the length scale of the coherence length $\xi$.

Experimental evidence for surface superconductivity has been found by dc transport\cite{Hempstead63,Smith68,Rothwarf67, Kirschenbaum75} or inductive measurements\cite{Strongin64} shortly after the
theoretical prediction.\cite{Saint-James63}
Later on, other methods such as ac-susceptibility and permeability measurements,\cite{Strongin64,Cruz69,Rollins70,Hopkins74} magnetization measurements,\cite{Schweitzer66,McEvoy69} surface impedance measurements\cite{Brunet74} and tunneling spectroscopy\cite{Strongin65} confirmed the existence of surface superconductivity when $\bm H$ was applied parallel to the surface.
The evolution of the resistance $R$ vs $H$, depending on the orientation of $\bm H$ relative to the surface was also investigated.\cite{Hempstead63}
While two different regions for bulk and surface superconductivity were clearly observed for fields parallel to the surface, no signature for surface superconductivity was observed when $\bm H$ was applied perpendicularly.
The in-plane-field dependence of the critical current $I_c(H)$ in the regime of surface superconductivity for $\bm H$ parallel to the bias current was described by Abrikosov\cite{Abrikosov65} and studied experimentally.\cite{Bellau66,Bellau67,Lowell69}
Park described theoretically the evolution of $I_c(H)$ in the state of surface superconductivity when the in-plane field $\bm H$ is applied perpendicular to the bias current flow\cite{Park65}.
He predicted an asymmetry in the critical surface current, resulting from the superposition of surface screening currents and external currents.
Such an asymmetry has not been observed experimentally yet.

Similar to surface superconductivity, localized superconductivity can also nucleate near the sample edge in a thin semi-infinite superconducting film, in a thin superconducting disk of very large diameter or around holes in a perpendicular magnetic field.\cite{White66,Bezryadin95,Bezryadin96,Berger02,Chibotaru05,Aladyshkin07}
It should be mentioned, that surface superconductivity and localized states at the sample edges in perpendicular field [called edge superconductivity (ES)] are qualitatively and quantitatively the same.
While surface superconductivity has been investigated in several  compounds like Pb-based alloys\cite{Strongin64,Schweitzer66,Kirschenbaum69}, Nb and Nb-based alloys,~\cite{Brunet74,Hopkins74,Schweitzer66,Kirschenbaum75}  polycrystalline MgB$_2$\cite{Tsindlekht06}, Pb\cite{Strongin64,Strongin65,Fischer68}, UPt$_3$ whiskers\cite{Keller96}, NbSe$_2$\cite{DAnna96}, experimental studies on ES in thin film structures are rare \cite{Stamopoulos04,Scola05}.
Recently, the first real space observation of ES was obtained by scanning tunneling microscopy on Pb thin film islands\cite{Ning09}.

Localized states do not only occur at sample boundaries but can also be induced by an inhomogeneous magnetic field as it appears e.g.~above domain walls in superconductor/ferromagnet hybrids.
This localized state is therefore called domain wall superconductivity (DWS)\cite{Buzdin03a,Aladyshkin03,Yang04b}.
Recently,  a Pb/BaFe$_{12}$O$_{19}$ superconductor/ferromagnet hybrid has been investigated by low-temperature scanning laser microscopy (LTSLM)  and the inhomogeneous current distribution of the sample in the DWS state has been visualized.\cite{Werner11}
LTSLM is therefore a valuable tool to visualize the redistribution of the current in the crossover from bulk to edge superconductivity.

In this paper we present our investigations on the evolution of edge superconductivity in plain and antidot Nb microbridges in perpendicular magnetic field.
Measurements of $R(H)$ were performed to compose an experimental phase diagram and to identify the regions of bulk and edge superconductivity.
Then we use LTSLM to visualize the current distribution at  the transition from the superconducting to the normal state in both bridges.
In addition, we used a time--dependent Ginzburg--Landau model to compare our experimental findings with theoretical predictions.

\section{Sample fabrication and experimental details}
\label{Section-II}

\begin{figure}[h]
\includegraphics[width=0.35\textwidth]{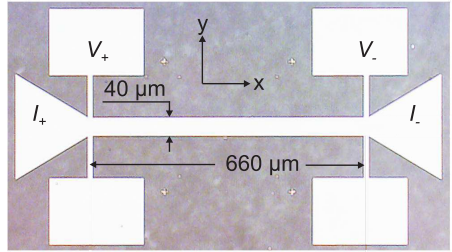}
\caption{Optical image of the plain $40\,\mu$m wide Nb bridge.
The contact pads used for $I$ and $V$ are indicated.}
\label{fig1}
\end{figure}

A Nb thin film with thickness $d=60\,$nm was deposited on a single crystal Al$_2$O$_3$ substrate (r-cut sapphire) at $T \approx 800^{\circ}\,$C using  magnetron sputtering.
Two Nb bridges with  width  $W=40\,\mu$m and  length $L=660\,\mu$m were patterned by e-beam lithography and reactive ion etching into a bridge geometry as shown in Fig.~\ref{fig1}.
One of these bridges was patterned with circular microholes [antidots (ADs) with 580\,nm diameter] in a triangular lattice with a period of $1.5\,\mu$m.

The samples were electrically characterized in a Helium cryostat at 4.2\,K$\le T\le$10\,K and $|H|\le$20\,kOe using a conventional four-terminal scheme (cf.~Fig.~\ref{fig1}).
For both investigated  Nb microbridges we found $T_{\mathrm{c0}}=$8.5\,K.
$\bm H$ was always applied along the $z$-direction, i.e.~${\bm H}=H\hat{\bm e}_z$ was perpendicular to the thin film surface and the applied bias current $I$.
We performed isothermal measurements of voltage $V(I)$ characteristics for different $T$ and $H$ values out of which we determined the dependence of the dc resistance $R=V/I$ on $H$.
The data presented below were obtained with $I$=1\,mA, unless stated otherwise.
This corresponds to a bias current density $J\equiv I/dW\approx 40\,$kA/cm$^2$.

To visualize the current distribution for different bias points in the $T-H$ phase diagram, we used LTSLM.\cite{Wagenknecht06,Wang09,Aladyshkin11,Werner11}
For imaging by LTSLM, the sample was mounted on a cold finger of a Helium flow cryostat, which is equipped with an optical window to enable irradiation of the sample in the $(x,y)$ plane by a focused laser beam with beam spot diameter $\sim 1.5 -2\,\mu$m.\cite{Wagenknecht06,Wang09}
The amplitude modulated laser beam (at frequency $f\approx 10\,$kHz) induces a local increase of temperature centered at the beam spot position $(x_0,y_0)$ in the sample.
During imaging, the Nb bridge is biased at a constant $I$, and the beam-induced change of voltage $\Delta V(x_0,y_0)$ is recorded by lock-in technique as a function of the beam coordinates $(x_0,y_0)$.
The LTSLM voltage signal can be interpreted as follows:
If the irradiated part of the sample was in the normal state with resistivity $\rho_n$, the laser beam induces a very small voltage signal $\Delta V\propto\partial \rho_n /\partial T$.
However, if the irradiated part of the bridge took part in the transfer of a substantial part of the superconducting currents, the beam-induced suppression of superconductivity might switch the whole sample from a low-resistive state to a high-resistive state.
Details of the LTSLM signal interpretation can be found in Refs.~[\onlinecite{Wagenknecht06,Wang09,Aladyshkin11,Werner11}].

\section{Results and discussion}

\subsection{Magnetoresistance data and Ginzburg-Landau simulations}

\begin{figure}[hbt]
\includegraphics[width=0.48\textwidth]{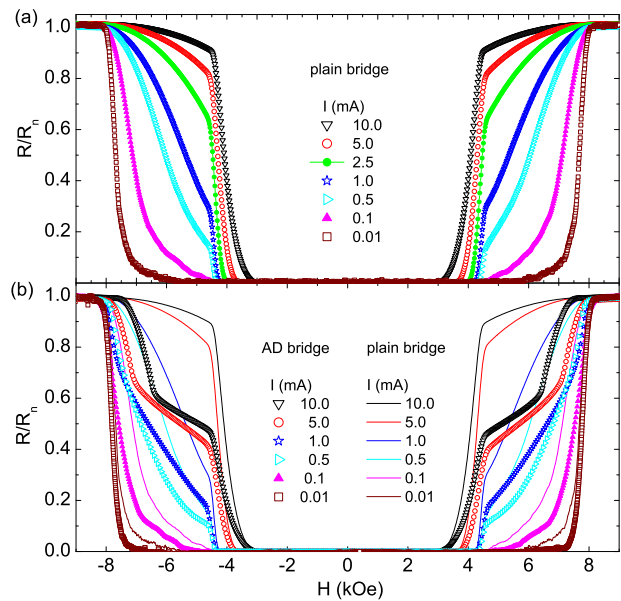}
\caption{(Color online) $R(H)$ curves (normalized to normal state resistance $R_n$) measured at $T$=4.2\,K with different bias currents $I$=0.01--10\,mA for (a) the plain and (b) the AD bridge. To facilitate the comparison, in (b) the data from (a) are shown again as thin lines.}
\label{fig2}
\end{figure}

Figure \ref{fig2} shows $R(H)$ measurements of the resistive transition at $T$=4.2\,K for different values of $I$ for the plain [Fig.~\ref{fig2}(a)] and the AD bridge [Fig.~\ref{fig2}(b)].
All curves are normalized to the normal state resistance $R_n$ at $H$=9.0\,kOe.
Except for the AD bridge at the highest current value of 10\,mA, all $R(H)$ curves reach $R_n$ at the same field value $|H|\approx 8\,$kOe.
However, we observe a pronounced dependence of the shape of the $R(H)$ curves on $I$, which we describe and discuss in the following.

For the plain bridge [cf.~Fig.~\ref{fig2}(a)], at the highest current value of 10\,mA, we observe with increasing $|H|$ an onset of dissipation (appearance of a finite $R$) at $\sim$3\,kOe.
Upon further increasing $|H|$, the slope $dR/d|H|$ steadily increases, yielding a rather steep $R(|H|)$ transition curve up to $\sim$0.9$R_n$.
At $\sim$0.9$R_n$ ($|H|\sim$4.4\,kOe) a kink in $R(|H|)$ appears, i.e.~with further increasing $|H|$, the slope $dR/d|H|$ is significantly reduced.
Upon reducing $I$, the field value where the kink appears stays almost constant; however, the resistance at the kink steadily decreases, and becomes zero for $I<$0.5\,mA, i.e.~the kink disappears.
Similar shapes of the $R(H)$ curves (including the above described kink) and their current dependence as shown in Fig.~\ref{fig2}(a) for the plain bridge have been found in [\onlinecite{Hempstead63,Kirschenbaum75}] when $\bm H$ was applied parallel to the sample surface.

The $R(H)$ measurements of the AD bridge [cf.~Fig.~\ref{fig2}(b)] show similar behavior upon variation of $I$ as compared to the plain bridge in the following sense:
Within the same range of (high) bias currents, the onset of dissipation appears almost at the same $H$ value (for the same value of $I$) as for the plain bridge.
Upon further increasing $|H|$, a similar steep transition (with slightly smaller slope as for the plain bridge) appears, up to the kink in $R(|H|)$, which is also present for the AD bridge within the same range of (high) bias currents.

However, we also observe distinct differences by comparing the AD and plain bridge: The resistance at the kink is lower for the AD bridge.
This deviation increases with increasing bias current.
Furthermore, for the two highest $I$ values the AD bridge shows a second kink in the $R(H)$ curve where the slope $dR/d|H|$ suddenly increases again with increasing $H$; this feature is absent at smaller $I$ and is not seen for the plain bridge for all values of $I$.
Finally, for the two lowest $I$ values, the onset of dissipation (upon increasing $|H|$) is shifted to larger $|H|$ values for the AD bridge, as compared to the plain bridge.

In the following, we present an interpretation of the $R(H)$ curves described above, starting with the discussion of the results obtained for the plain bridge.
At the highest $I$=10\,mA, upon increasing the external magnetic field from $H=0$, vortices will enter the sample when $H$ is larger than the field of first vortex entry, which is rather small for thin-film structures in perpendicular magnetic field.
The onset of energy dissipation can then be attributed to the onset of motion of vortices, when the bias current density $J$ exceeds the depinning current density $J_{dpin}$ at a given $T$ and $H$.
In this case, upon further increasing $H$, the flux flow resistance will strongly increase, i.e.~, the rather large slope $dR/dH$ should correspond to the bias-current-stimulated motion of the vortex lattice in the presence of a strong pinning potential.
The kink in the $R(H)$ curve where the slope $dR/dH$ substantially decreases (upon increasing $H$), can be assigned to the transition from the resistive flux-flow regime to a resistive regime with fully suppressed bulk superconductivity and surviving ES at $H_{c2}$ (and above).
This interpretation is the same as given in  [\onlinecite{Kirschenbaum75}] (for surface superconductivity with $\bm H$ parallel to the sample surface).
However, in contrast to our observation, a more gradual transition to $R_n$ already at $H_{c2}$ without any kinks and no signature of ES was observed in [\onlinecite{Hempstead63,Kirschenbaum75}] when $\bm H$ was applied perpendicular to the sample surface.

We would like to emphasize that the position of the kink should be close to the upper critical field $H_{c2}$ but not identical to it, since the destruction of bulk superconductivity is a
thermodynamical property of a material, but the kink can be observed only under strong non-equilibrium conditions upon the bias current injection.
Still, below we use the field value where the kink appears as the experimentally determined $H_{c2}$ value.

Obviously, in our case the edge states form continuous channels with enhanced conductivity, which reduce the overall resistance to a value below $R_n$.
The observed reduction of the resistance at the kink feature in $R(H)$ with decreasing $I$ can be explained by the strengthening of ES upon decreasing $I$, until at small enough currents the injected bias current flows entirely as a dissipationless supercurrent along the edge channels at $H=H_{c2}$, leading to a disappearance of the kink feature.

As described above, the full normal resistance $R_n$ is reached (for all $I$ values) at the same field, which we now associate with the upper critical field $H_{c3}\approx 1.7H_{c2}$ for ES.
An analysis of the $T$ dependence of $H_{c2}$ and $H_{c3}$ will be presented in Sec.~\ref{H-T-diagram}.

In the AD bridge, the holes lead to additional ``edges'' in the sample interior, which
results in a higher volume fraction of ES and more effective pinning.
This explains the lower $R$ value (as compared to the plain bridge) at the kink when bulk superconductivity becomes suppressed at $H_{c2}$.
The origin of the second kink at $H_{c2}$$<$$|H|$$<$$H_{c3}$, developing at rather large bias current [Fig.~\ref{fig2}(b)], might be associated with a slightly reduced $H_{c3}$ value at the AD edges, as compared to the edges of the bridge, due to the different edge geometry.
However, further investigations are required to provide a more conclusive explanation on this feature.
Similarly, we cannot yet provide an explanation for the observed shift of the onset of dissipation to larger $H$, for the AD bridge (as compared to the plain bridge) for the lowest values of $I$.

\begin{figure}[t!]
\includegraphics[width=0.48\textwidth]{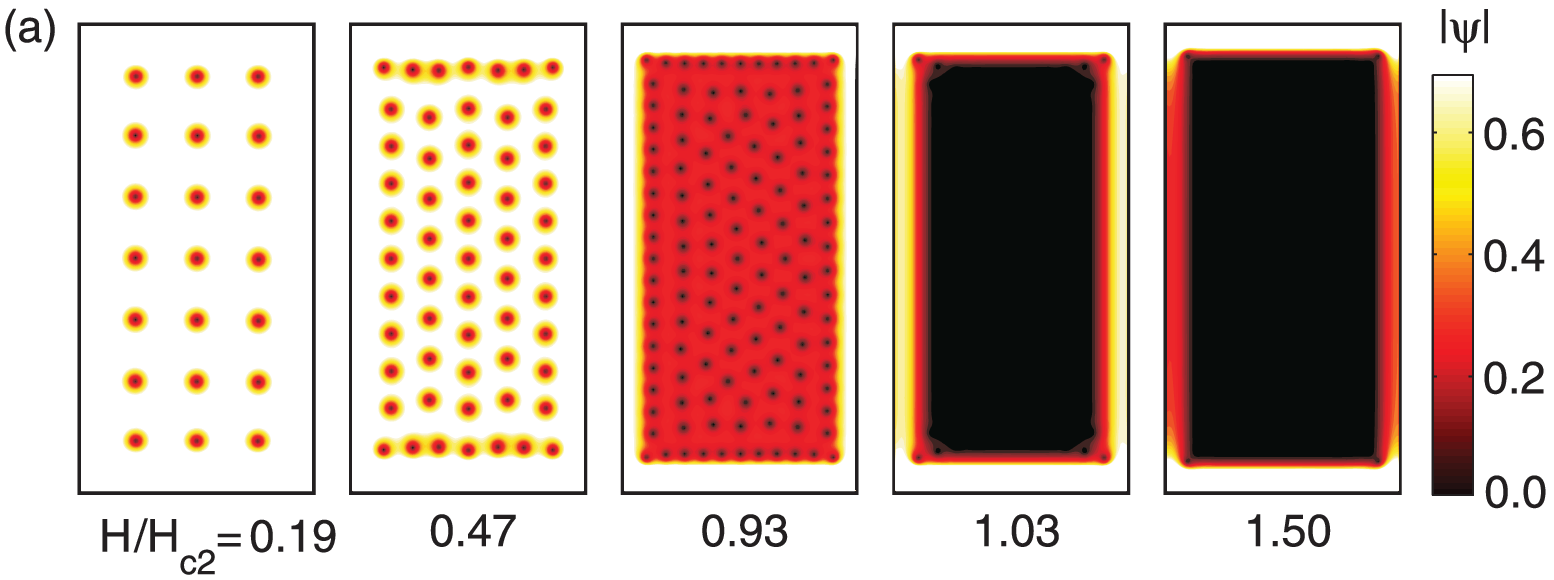}
\includegraphics[width=0.48\textwidth]{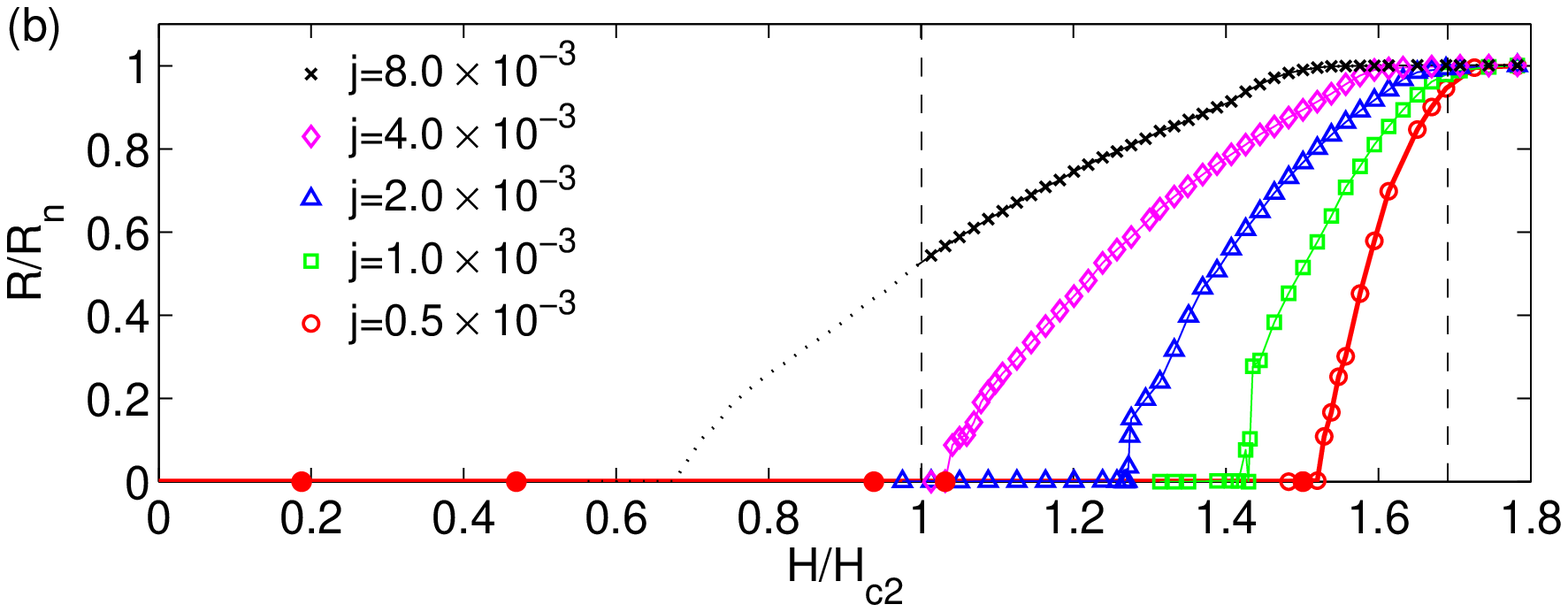}
\caption{ (Color online) Numerical GL-simulation results for a superconducting rectangular thin film ($W=30\,\xi_0$, $L=60\,\xi_0$) biased at normalized current density $j$ at variable magnetic field $H$ and $T/T_c$=0.47.
(a) Spatial distribution of the modulus of the normalized order parameter wave function $|\psi(x,y)|$ with $j$=0.5$\times 10^{-3}$ [lowest value in (b)].
The five panels show simulation results for different values of $H/H_{c2}$; $j$ is flowing from top to bottom.
(b) $R/R_n$ vs $H/H_{c2}$ for different $j$.
Two vertical dashed lines depict the upper critical field $H_{c2}$ and the critical field of ES $H_{c3}=1.695\,H_{c2}$.}
\label{fig3}
\end{figure}

To compare the experimental results with theoretical calculations based on the Ginzburg-Landau (GL) model described in the Appendix, we calculated for a rectangular plain superconducting thin film ($W= 30\, \xi_0$, $L=60\, \xi_0$; $\xi_0$ is the GL coherence length at $T=0$) the spatial distribution and time dependence of the normalized order parameter (OP) wave function $\psi(x,y,t)$ and the voltage drop $V(t)$ along the rectangle for different values of $H$ and normalized bias current density $j$ at a reduced temperature $T/T_c$=0.47 (corresponds to $T$=4.2\,K for Nb with $T_c$=9\,K).
We want to note, that the real dimensions of the investigated sample exceed considerably the dimensions used in our modeling. 
Nevertheless, the model correctly captures the essential physics behind the discussed effects for $H>H_{c2}$.
Figure \ref{fig3}(a) shows the spatial distribution $|\psi(x,y)|$
for a rather small value of $j$=$5\times 10^{-4}$ for five different values of $H/H_{c2}$ from 0.19 to 1.50.
We note that the chosen value for $j$ is several orders of magnitude below the GL depairing current density $j_{\mathrm{GL}}$=0.386 [cf.~the appendix] at $T$=0.
In all cases, the OP distributions are time-independent (''stationary case``) corresponding to zero resistance.
At $|H|<H_{c2}$, a regular vortex structure appears and the density of vortices increases with increasing $H$.
%
%
However, even when bulk superconductivity is depleted at $|H|>H_{c2}$, superconducting channels with finite and time-independent $|\psi|$ running along the edges of the rectangle are still present and can provide a non-dissipative current transfer.
If $H$ is further increased, the superconductor turns to a non-stationary regime with finite resistance and reaches its normal value at the upper critical field for ES at $H= H_{c3}$.
The calculated $R/R_n$ vs $H/H_{c2}$ curves for different $j$ are shown in Fig.~\ref{fig3}(b).
The numerical simulations reproduce the shift of the curves to smaller $H$ and the decrease
of the slope $dR/dH$ in the interval $H_{c2}<|H|<H_{c3}$ as $j$ increases.
This is qualitatively the same as observed experimentally in Fig.~\ref{fig2}.
However, for large enough $j$ (finite $R$ at $H_{c2}$) our model is unable to describe the kink in $R(H)$ close to $H_{c2}$ and the disappearance of $R$ for $H<H_{c2}$, since bulk pinning was not taken into account [cf.~curve for $j=8\times 10^{-3}$ in Fig.~\ref{fig3}(b)]).

\subsection{Superconducting phase diagram for the plain bridge}
\label{H-T-diagram}

\begin{figure}[htb]
\includegraphics[width=0.48\textwidth]{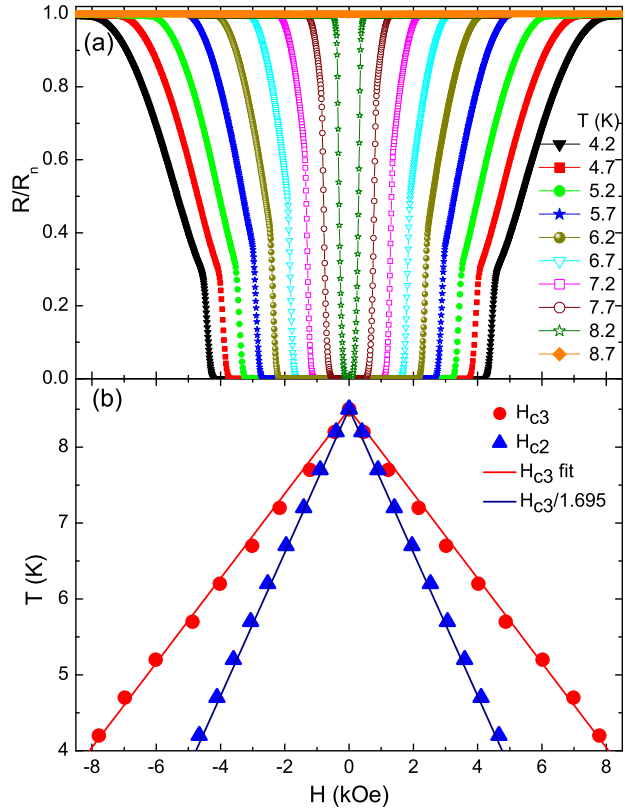}
\caption{(Color online) (a) $R(H)$ curves (normalized to normal state resistance $R_n$) of the plain bridge for $T$=4.2--8.7\,K  (from outside to inside).
(b) $H-T$-phase diagram of the plain bridge.
Data points for $H_{c3}$ (dots) and $H_{c2}$ (triangles) are deduced from $R(H)$ curves in (a).
The lines are the transition lines for ES, which were fitted to the data points with Eq.~(\ref{eq:Hc3}) and $T_{c0}=8.5\,$K, and the calculated transition lines for $H_{c2}$ using the relation $H_{c3} = 1.695\,H_{c2}$}
\label{fig4}
\end{figure}

Figure~\ref{fig4}(a) shows the results of the $R(H)$ measurements for $T$=4.2--8.7\,K.
With increasing $T$, the deviation from $R=0$ and the kink, both shift to smaller $H$ values, and  the resistance at the kink shifts to a higher $R/R_n$ ratio, while the change in the slope $dR/dH$ at the kink becomes less pronounced.
In order to experimentally determine  $H_{c2}$ and  $H_{c3}$ we use the field value at the kink and a criterion of 0.98\,$R_n$, respectively.
%
%
%
The determined transition lines for $H_{c2}(T)$ and $H_{c3}(T)$ for the above mentioned criteria are shown in Fig.~\ref{fig4}(b).
The experimental transition line for $H_{c3}(T)$ can be fitted with Eq.~(\ref{eq:Hc3}) and $T_{c0}=8.5\,$K which extrapolates to $H_{c3}^{(0)}$=15.3\,kOe.
Plotting the transition line for $H_{c2}(T)$ with the relation $H_{c3}(T) = 1.695 \,H_{c2}(T)$, we find that the experimentally determined $H_{c2}$ values are close to the calculated transition line for bulk superconductivity.
This result gives convincing evidence that depending on $T$, $I$ and $H$ (perpendicular to the sample surface), our sample can be either in the state with developed bulk superconductivity and pinned vortex lattice ($H<H_{c2}$), or in the resistive state, controlled by ES ($H_{c2}<H<H_{c3}$).

\subsection{Visualization of the current distribution by LTSLM}

\begin{figure*}[thb]
\includegraphics[width=1\textwidth]{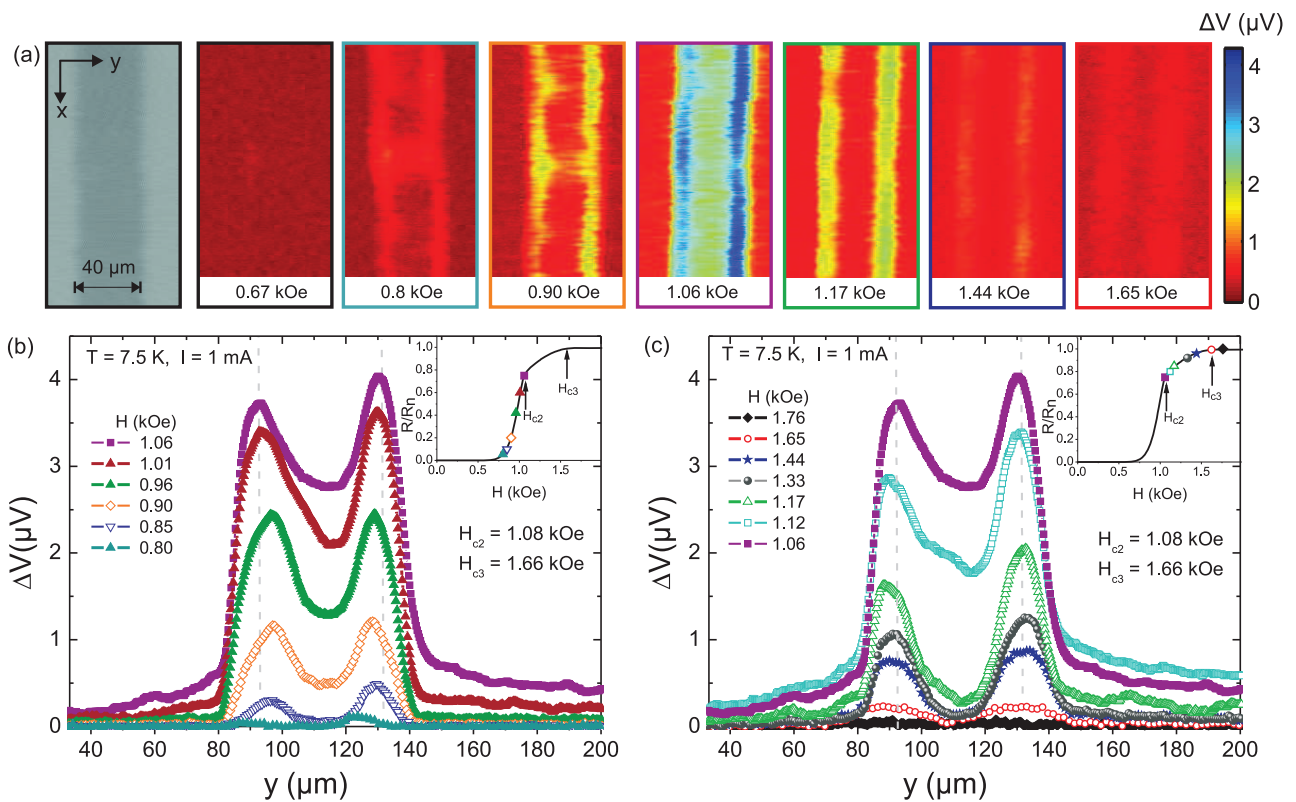}
\caption{(Color online) LTSLM signals from plain Nb bridge at $T$=7.5\,K.
(a) Optical image (left) and voltage images $\Delta V(x,y)$ for different $H$;
(b) linescans $\Delta V(y)$ across the bridge for $H \leq H_{c2}$ and (c) for $H \geq H_{c2}$.
The insets in (b) and (c) show the $R(H)$ curve with corresponding bias points for the linescans.
Vertical dashed lines in (b) and (c) indicate position of the edges of the bridge}
\label{fig5}
\end{figure*}

We used LTSLM to visualize the current distribution in the Nb bridges during the transition from bulk superconductivity to the normal state.
As the maximum $H$ was limited to $\sim 2\,$kOe in this setup, the LTSLM measurements were performed at rather high $T$ values, $T=7.0 - 7.5\,K$.

Figure \ref{fig5}(a) shows an $H$-series of  beam-induced voltage images, $\Delta V(x,y)$, at $T$=7.5\,K for various superconducting states of the plain Nb bridge, oriented vertically in all these images [cf.~optical image (left panel) in Fig.~\ref{fig5}(a)].
For a more quantitative analysis, we show an $H$-series of linescans, $\Delta V(y)$, across the bridge in Fig.~\ref{fig5}(b) and (c).
The insets in Fig.~\ref{fig5}(b) and (c) show the corresponding  $R(H)$ curve, from which we estimate $H_{c2}\approx$1.1\,kOe  and $H_{c3}\approx$1.8\,kOe.
At $H = 0.67\,$kOe in Fig.~\ref{fig5}(a), the LTSLM signal is zero, which means that the beam-induced perturbation is not strong enough to suppress superconductivity and to induce a voltage signal.
Upon increasing $H$, the  first signal appears at $H\approx 0.8\,$kOe which corresponds to the onset of the resistive transition [see inset in Fig.~\ref{fig5}(b)].
With further increasing $H$, the signal at the edges is enhanced, but also a signal from the inner part of the bridge appears.
The latter can be attributed to the depletion of bulk superconductivity with increasing $H$ (below $H_{c2}$), which leads to an increasing voltage response to the perturbation by the laser beam with a maximum beam-induced voltage signal at $H=1.06\,$kOe, which is very close to the estimated $H_{c2}$ value.
The pronounced edge signal below $H_{c2}$ can be explained by the suppression of the edge barrier for vortex entry/exit by the laser spot.
%
%
Hence one can expect that irradiation at the edges of the bridge should strongly affect the vortex pattern and the resulting current distribution.
In contrast, laser irradiation of the interior of the bridge does not change the energy barrier and the modification is probably less pronounced and the signal in the interior is much smaller.

Fig.~\ref{fig5}(c) shows linescans for $H \geq H_{c2}$.
For fields larger than $H_{c2}$, the beam-induced voltage in the center of the bridge drops almost to zero while large peaks are still observed at the edges of the bridge.
This apparently reflects the fact that above  $H_{c2}$, the bulk is no longer superconducting and therefore does not lead to a voltage signal, while the edges still contribute to a strong LTSLM signal due to ES.
The rather large width of these edge peaks in the state of ES can be explained by the fact that the edge states are not only perturbed when the laser beam spot is centered right at the edges, but also when the tail of the beam-induced heat distribution leads to a suppression of the edge states when the beam is centered slightly off the edges.
A further increase in $H$ leads to a gradual decrease of the edge peaks which finally disappear at $H = 1.76\,$kOe which is close to $H_{c3}$.
Above $H_{c3}$, the sample is completely in the normal state and the effect
of the laser beam on the resistive state is negligible.

In summary, the linescan series in Fig.~\ref{fig5}(c) indicate, that above $H_{c2}$, the dominant part of the current is flowing at the edges of the sample.
Thus, LTSLM seems to be a capable to visualize the ES states and to identify the different regimes in the $R(H)$ curves for the plain bridge.

\begin{figure*}[htb]
\includegraphics[width=0.98\textwidth]{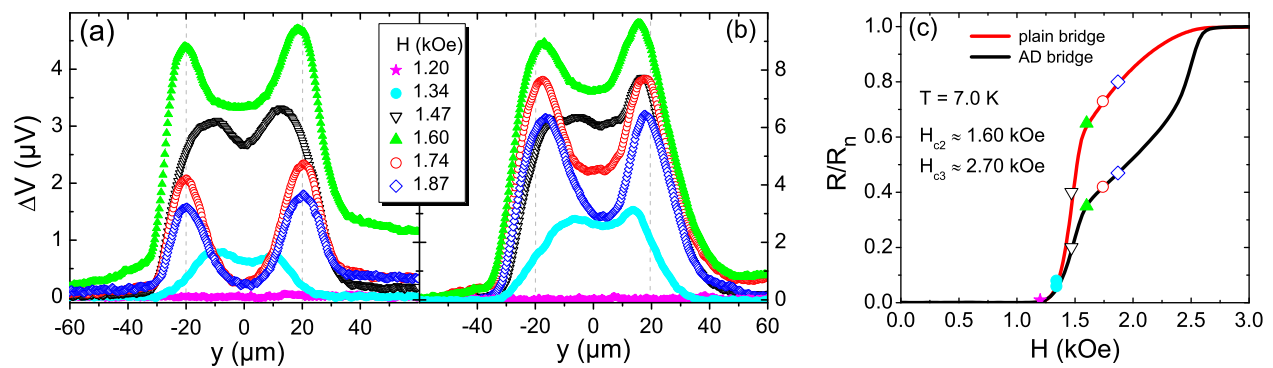}
\caption{(Color online) LTSLM linescans $\Delta V(y)$ for variable $H$ at $T$=7.0\,K across (a) the plain and (b) the AD bridge; note the different $\Delta V$ scale{} in (a) and (b).
Vertical dashed lines indicate position of the edges of the bridge
(c) Corresponding $R(H)$ curves with bias points for the linescans in (a,b).}
\label{fig6}
\end{figure*}

For comparison, we show a linescan series (variable $H$) for the plain [Fig.~\ref{fig6}(a)] and AD bridge [Fig.~\ref{fig6}(b)] at $T=7.0\,$K.
The corresponding $R(H)$ curves with the bias points of the linescans are shown in Fig.~\ref{fig6}(c).
The  $H_{c2}$ value for this temperature is $\sim 1.6\,$kOe.
We note, that the beam-induced signal of the AD bridge is higher than for the plain bridge, which we ascribe to the higher current density in the AD bridge due to its reduced cross section  because of the holes.
For the lowest field value, $H=1.20\,$kOe, the beam-induced heating of the laser has no effect, while  at $H=1.34\,$kOe the whole cross section of both bridges leads to a LTSLM signal.
As $H$ increases further, the LTSLM signal from the edges becomes larger than the signal from the interior and the overall signal increases up to $H = H_{c2}=1.60\,$kOe.
For the plain bridge the overall signal gets strongly reduced above $H_{c2}$, and the signal from the central part of the bridge almost vanishes.
The key difference between the plain and AD bridge is that for the latter sample the voltage signal gets much less reduced and the whole cross section of the AD bridge gives a measurable signal.
This means that the current is distributed across the entire width of the bridge even for $H>H_{c2}$.
This observation is consistent with the $R(H)$ measurements shown in Fig.~\ref{fig6}(c), where the additional edges inside the AD bridge lead to a different shape in $R(H)$ and a lower $R$ for any value of $H$ within the interval $H_{c2}\lapprox H \lapprox H_{c3}$.

\subsection{Bias-current-induced asymmetry: LTSLM response and Ginzburg-Landau simulations}

According to Fig.~\ref{fig5}(b,c), the LTSLM signal $\Delta V(y)$ is asymmetric with respect to the bridge center (axis $y$=0) for several $H$ values around $H_{c2}$, i.e. the right maximum is slightly higher than the left one.
This asymmetry in the beam-induced voltage response can be explained  by an asymmetry in the supercurrent density distribution $j_{s,x}(y)$ close to the left and right edge.
Based on the time dependent GL model\cite{Comment-GLDD}, we calculate the time-averaged quantities for the OP distribution $\langle|\psi|^2\rangle(y$) and the $x$--components of the superfluid current density $\langle j_{s,x}\rangle(y$) and the normal current density $\langle j_{n,x}\rangle(y$).

\begin{figure}[h]
\includegraphics[width=0.48\textwidth]{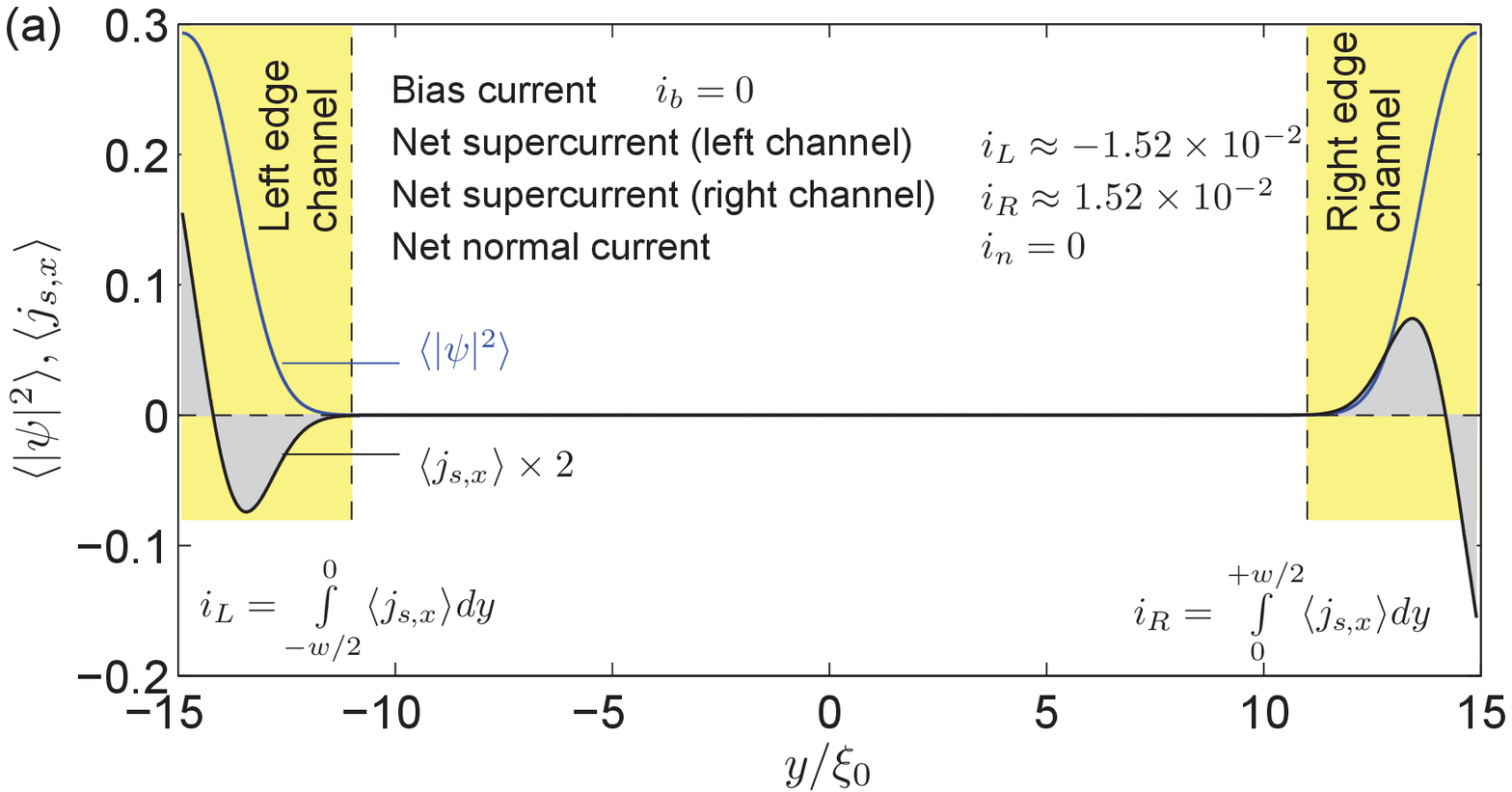}
\includegraphics[width=0.48\textwidth]{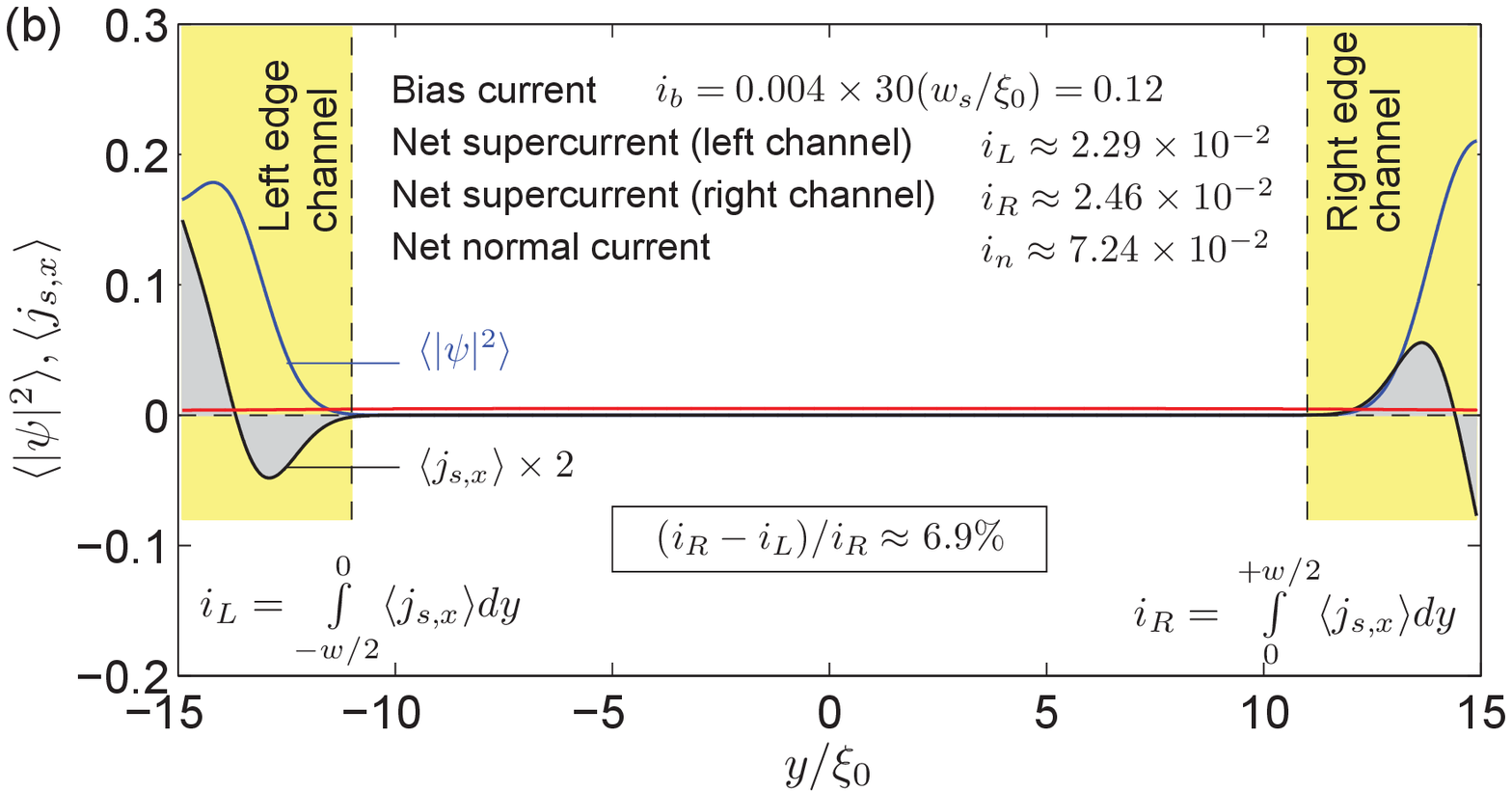}
\includegraphics[width=0.48\textwidth]{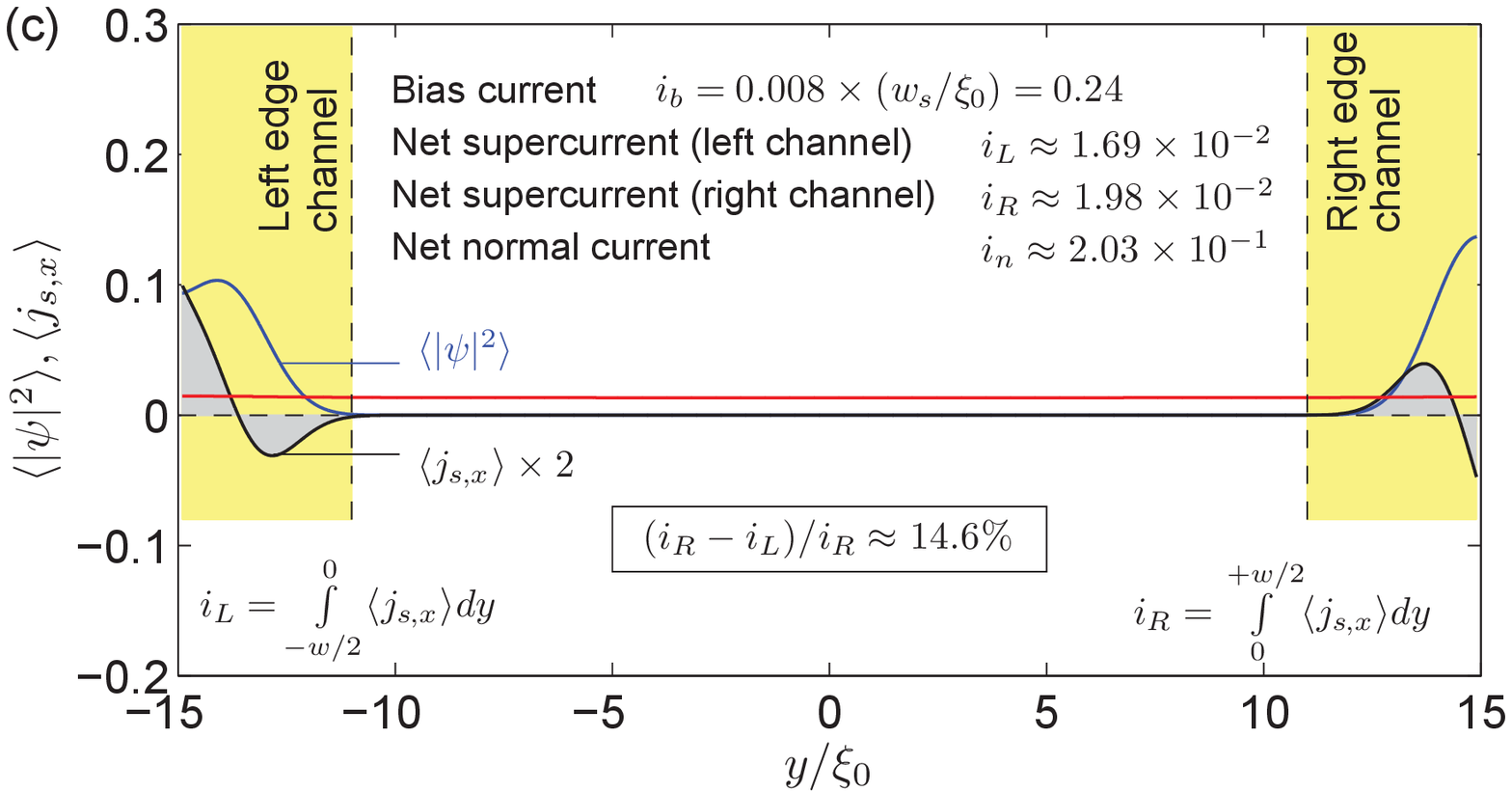}
\caption{(Color online) Time-averaged normalized OP wave function $\langle|\psi|^2\rangle(y/\xi_0)$ and $x$--components of the normalized superfluid current density $\langle j_{s,x}\rangle(y/\xi_0)$ and normalized normal current density $\langle j_{n,x}\rangle(y/\xi_0)$, calculated for a rectangular thin film ($W=30\,\xi_0$, $L=60\,\xi_0$; cf.~Fig.~\ref{fig3}) at $T/T_c$=0.47 and $H=1.3H_{c2}$.
The three graphs differ in the normalized bias current (a) $i_b$=0, (b) $i_b$=0.12 and (c) $i_b$=0.24.
$i_L$ and $i_R$ denote the integrals of $\langle j_{s,x}\rangle$ (shaded areas) over the left and right edge channels, respectively.}
\label{fig7}
\end{figure}

Figure \ref{fig7} shows results of such calculations for $H=1.3\,H_{c2}$ and $T/T_{c0}$=0.47, which were obtained for zero bias current density $j=j_{s,x}+j_{n,x}$ [Fig.~\ref{fig7}(a)], for $j$ close to the critical current density at $H=H_{c2}$ [Fig.~\ref{fig7}(b)] and for $j$ which is larger than the critical current density for ES  within the entire field range $H_{c2}<H<H_{c3}$ [Fig.~\ref{fig7}(c)].

According to our calculations, even in the resistive ES state, there is a finite superfluid flow localized within the ES channels.
These supercurrents are circulating in opposite direction within each of the two edge channels, which is due to the applied magnetic field $H$.

For further analysis, we determined the net currents $i_L$, $i_R$ and $i_n$.
Here, $i_L$ and $i_R$ are the integrals of $j_{s,x}$ across the left and right edge channel, respectively (shaded areas in Fig.~\ref{fig7}); $i_n$ is the integral of $j_{n,x}$ across the entire width of the rectangular film.
Hence, for the normalized bias current $i_b\equiv j\frac{W}{\xi_0}$ we have $i_b=i_L+i_R+i_n$.

For $j=0$ ($i_b=0$) [cf.~Fig.\ref{fig7}(a)], the net currents $i_L$ and $i_R$ in the right and left edge channel have the same finite amplitude, but differ in sign, and $i_n=0$.
%
%
For $j > 0$ ($i_b > 0$) [cf.~Fig.\ref{fig7}(b,c)] the steady-state distribution of the superconducting parameters differs from the case $j=0$.
Now, $i_L$ and $i_R$ do have the same (positive) sign, but different amplitudes.
Thus, analyzing only the large-scale details in the supercurrent distribution (spatially averaged over length scales much larger than the coherence length $\xi_0$), one can think in terms of a combination of two parallel currents flowing along the sample edges with different amplitudes depending both on $I$ and $H$ direction.
It should be noted that a very similar situation -- the asymmetry of the critical current density -- was described by Park [\onlinecite{Park65}] within a stationary
Ginzburg-Landau model.
Since the mentioned asymmetry results from the superposition of the bias current $I$ and the currents induced by the applied $H$ field, the asymmetry can therefore being changed either by changing the current direction or the sign of $H$.

\begin{figure}[h]
\includegraphics[width=0.48\textwidth]{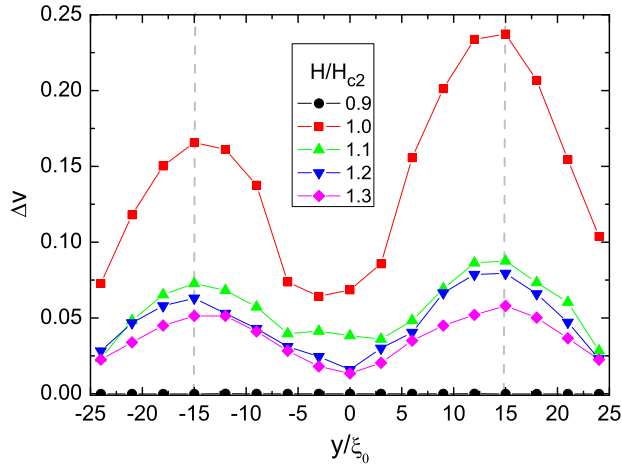}
\caption{(Color online) Calculated normalized LTSLM beam-induced  voltage $\Delta v$=$\bar{v}_\mathrm{on}$-$\bar{v}_\mathrm{off}$ vs $y/\xi_0$ across a rectangular thin film ($W=30\,\xi_0$, $L=60\,\xi_0$; cf.~Figs.~\ref{fig3} and \ref{fig7}) for different values of $H$$/H_{c2}$ at $T/T_c$=0.47 and $j$=4$\times 10^{-3}$.
The vertical dashed lines indicate the position of the edges.}
\label{fig8}
\end{figure}

In order to prove, whether the asymmetry of the LTSLM signal can be related to the bias-current-induced asymmetry, we calculated the normalized beam-induced voltage $\Delta v(y)$, i.e., linescans across a rectangular superconducting thin film with the geometry as in Fig.~\ref{fig3} and Fig.~\ref{fig7}, biased at $j=4\times 10^{-3}$.
Details of the calculation can be found in the Appendix.
Assuming a Gaussian shape of the laser-beam-induced increase in $T$ with a maximum amplitude $\Delta T$ and a full width half maximum of $\sigma =7 \xi_0$, we obtain the linescan series for different values of $H/H_{c2}$ shown in Fig.~\ref{fig8}.
These simulations clearly show that the voltage signal has maxima near the left and right edges, and that their amplitudes are different, with this asymmetry being most pronounced at $H=H_{c2}$.
This is in nice agreement with experimental LTSLM results.

\begin{figure}[h]
\includegraphics[width=0.49\textwidth]{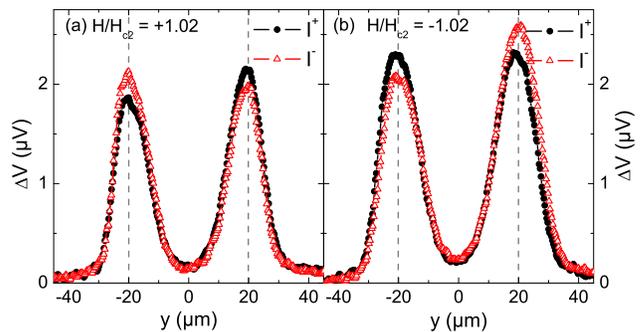}
\caption{(Color online) LTSLM linescans $\Delta V(y)$ across the plain Nb bridge ($T$=7.2\,K, $H_{c2}$=1.37\,kOe, $|H|$=1.40\,kOe, $|I|$=1\,mA) for different sign of $I$ and (a) negative $H$ and (b) positive $H$.
Vertical dashed lines indicate position of the edges of the bridge}
\label{fig9}
\end{figure}

To proof experimentally, that the asymmetry depends on sign of $H$ and $I$, we performed a series of LTSLM linescans on the plain Nb bridge.
The reversal of the asymmetry of the measured LTSLM signal upon the inversion of the $I$ and $H$ signs is illustrated in Fig.~\ref{fig9}(a) and (b).
We find that the right peak is larger for $I>0$ while the left peak is larger for $I<0$ and vice versa.
The slightly larger amplitudes of the peaks in Fig.~\ref{fig9}(b) are probably due to the residual field (in the 10\,Oe range) in the cryostat at the sample position.
To the best of our knowledge, this is the first direct experimental verification of an asymmetry in the current density in the ES state, as predicted by Park for surface superconductivity.

\section{Conclusion}

In this paper we studied experimentally and numerically the peculiarities of the resistive transition in thin-film Nb microbridges with and without antidots (ADs) in perpendicular magnetic field $H$.
From integral $R(H)$ measurements we find that the transition from bulk to edge superconductivity (ES), and finally to the full normal state, can be identified by a pronounced change in slope $dR/dH$, which however strongly depends on the applied bias current density.
The additional edges induced by the holes in the AD bridge lead to a different shape of the $R(H)$ curves as compared to the plain bridge.
The ES state as well as the evolution of superconductivity upon sweeping $H$ was imaged by low-temperature scanning laser microscopy (LTSLM).
For the ES state, LTSLM revealed an asymmetry in the currents flowing along the left and right edges, depending on the relative direction of applied current and external field, as proposed long time ago \cite{Park65}.
Our calculations based on the time--dependent Ginzburg--Landau theory confirm essential features of the experimental results.

\section{Acknowledgment}
This work was supported by the Russian Fund for Basic Research, RAS under the Program ''Quantum physics of condensed matter``, Russian Agency of Education under the Federal Target Program ''Scientific and educational personnel of innovative Russia in 2009--2013``, Deutsche Forschungsgemeinschaft (DFG) via grant no. KO 1303/8-1.
R. Werner acknowledges support by the Cusanuswerk, Bisch\"{o}fliche Studienf\"{o}rderung, D. Bothner acknowledges support by the Evangelisches Studienwerk Villigst e.V. and M. Kemmler acknowledges support by the Carl-Zeiss Stiftung.
The authors thank A.~I.~Buzdin for valuable discussions.


\section{Appendix}


In order to describe the general properties of the resistive state in a mesoscopic superconducting thin film sample and to compare them with experiment, we use a simple time--dependent Ginzburg--Landau (TDGL) model.\cite{IvlevKopnin84}
For simplicity we assume that the effect of the superfluid currents on the magnetic field distribution is negligible and consider the internal magnetic field $B$ equal to the external magnetic field $H$ (perpendicular to the thin film plane). This assumptions seems to be valid for the following two cases;  
(i) for mesoscopic thin-film superconductors with lateral dimensions smaller than the effective magnetic penetration depth $\Lambda = \lambda^2/d$ ($\lambda_L$ is the London penetration depth, $d$ is the thickness);
(ii) for superconductors for large $H$ and/or $T$ (i.e. close to the phase transition line), when the superfluid density tends to zero. 
Then the TDGL equations take the form
    \begin{eqnarray}
    u \left(\frac{\partial}{\partial t}+ i\varphi \right) \psi = \tau\,\left(\psi-|\psi|^2\psi\right) + \left(\nabla + i{\bf A}\right)^2\psi,
    \label{eq:GL-1}\\
    \tau = 1 - T({\bf r})/T_{c0}, \qquad\qquad\qquad\quad
    \label{eq:GL-2}\\
    \nabla^2 \varphi = {\rm div}\, {\bm j}_s, \,{\bm j}_s = -\frac{i}{2} \tau\, \Big\{\psi^*\left(\nabla + i{\bf A}\right)\psi - \mbox{c.c.}\Big\},
    \label{eq:GL-3}
    \end{eqnarray}
where $\psi$ is the normalized order parameter (OP), $\varphi$ is the dimensionless electrical potential, ${\bm A}$ is the vector potential [${\rm rot\,}{\bm A}=H\,\hat{\bm e}_z$], $T({\bm r})$ is local temperature (potentially position--dependent), ${\bm j}_s$ is the density of the supercurrent, $u$ is the rate of the OP relaxation, c.c. stands for complex conjugate.
We use the following units: $m^*\sigma_n \beta/(2e^2\tilde{\alpha})$ for time, the coherence length $\xi_0$ at temperature $T=0$ for distances, $\Phi_0/(2\pi\xi_0)$ for the vector potential, $\hbar e |\tilde{\alpha}|/(m^*\sigma_n \beta)$ for the electrical potential,
%
%
and $4e\tilde{\alpha}^2\xi_0/(\hbar\beta)$ for the current density, where $\alpha=-\tilde{\alpha}\,\tau$ and $\beta$ are the conventional parameters of the GL expansion, $e$ and $m^*$ are charge and the effective mass of carriers, $\sigma_n$ is the normal state conductivity.
In these units the Ginzburg--Landau deparing current density at $T=0$ is equal to 0.386.
We apply the boundary conditions in the following form
    \begin{eqnarray}
    \left(\frac{\partial}{\partial \bm n} + i A_n\right)_{\Gamma} \psi = 0,\quad \left(\frac{\partial \varphi}{\partial \bm n} \right)_{\Gamma} = j_{ext},
    \label{eq:BoundCond-1}
    \end{eqnarray}
where $\bm n$ is the normal vector to the sample's boundary $\Gamma$,
$j_{ext}$ is the normal component of the inward (outward) flow of the bias current density $\bm j$ (with $|\bm j|\equiv j$).
We do not consider bulk pinning, since the number of additionally required parameters (describing the spatial distribution of pinning sites and their pinning strength) would be too large.

\begin{figure}[htb]
\includegraphics[width=6cm]{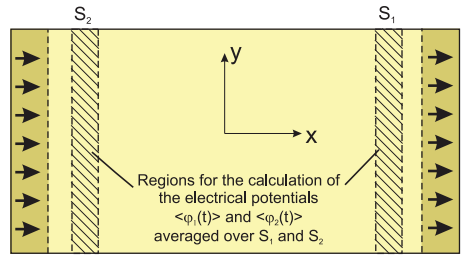}
\caption{(Color online) Schematic drawing (top view) of rectangular superconducting thin film bridge considered for GL simulations.
Arrows indicate injection and extraction of the bias current.
Shaded areas S$_1$ and S$_2$ are virtual electrodes.}
\label{fig10}
\end{figure}

%
We calculate\cite{Comment-GLDD} the instant value of the normalized voltage drop $v(t)=\langle\varphi_1(t) \rangle - \langle \varphi_2(t) \rangle $ and analyze the dependence of $v(t)$ on $H$ and $j_{ext}$.
Here
\begin{equation}
\langle \varphi_{i}(t) \rangle=\frac{1}{S_{i}} \int\int_{S_{i}} \varphi_{i}(x,y,t)\,dxdy
\label{eq:avg-varphi}
\end{equation}
is the time-dependent electrical potential averaged over the region $S_i$ ("virtual electrodes", $i=\{1,2\}$).
These regions have the same width as the sample width and they are shifted from the physical edges towards the sample interior (see Fig.~\ref{fig10}) for eliminating the effect of the sample edges.
In addition we formally consider an inhomogeneous sample, containing two areas at the left and right edges [cf.~Fig.~\ref{fig10}] with a critical temperature $T_{c1}$ (at $H=0$) and upper critical field $H_{c2,1}^0$ (at $T=0$) exceeding considerably $T_{c0}$ and $H_{c2}^0$ in the rest of the sample.
The reason for that is a pure technical one. This approach guarantees that the injected \textit{normal} current $i_b$ is fully converted into a supercurrent within these enhanced superconducting areas at any temperature and any value of $H$.
%

For the stationary regime all the calculated parameters, after transient processes induced by changes in the external parameters, tend to their
time--independent values, pointing out to the absence of energy dissipation for the established state and $R\to 0$.
For larger $T$, $H$ or $i_b$ the relaxation to the stationary case becomes impossible and all parameters oscillate in time.
Calculating the mean normalized voltage drop $\bar{v}$, averaged over a very large time interval (including up to $10^2$ of the voltage oscillations), one can determine the normalized beam--induced LTSLM voltage signal $\Delta v=\bar{v}_\mathrm{on}-\bar{v}_\mathrm{off}$, where $\bar{v}_\mathrm{on}$ and $\bar{v}_\mathrm{off}$ are the time averaged normalized voltage signals if the laser beam is on or off, respectively.

The effect of the focused laser beam can be treated as a quasistatic perturbation of the superconducting properties of the bridge, since the time scales of this perturbation are much longer than the GL time constant.
In the most simple form this perturbation can be modelled as a Gaussian--like increase in local temperature in Eq.~(\ref{eq:GL-2}):
\begin{equation}
T({\bf r})= T_0 + \Delta T\cdot e^{[-(x-x_0)^2 - (y-y_0)^2]/\sigma^2}
\label{eq:Gauss}.
\end{equation}
Here, $T_0$ is the sample temperature if the laser beam is off or far from the beam spot centered at $(x_0,y_0)$, $\Delta T$ is the amplitude of the local heating, depending on the beam intensity and on the rate of heat dissipation due to the thermal conductivity of the superconducting film and the substrate and on the thermal boundary resistance between the film and the substrate; $\sigma$ is the full width half maximum of the beam--induced temperature profile.\cite{Gross94}



\begin{thebibliography}{44}
\bibitem[{Com()}]{Comment-GLDD}
\bibinfo{note}{The described simulations were performed using the
  Windows--oriented solver GLDD, developed in the Institute for Physics of
  Microstructures RAS}.
\end{thebibliography}
\end{document}